\documentclass[aps,pre,groupedaddress,showpacs,preprint]{revtex4}
\usepackage{graphicx}% Include figure files
\usepackage[dvips]{epsfig}% Include figure files
\usepackage{dcolumn}% Align table columns on decimal point
\usepackage{subfigure}%
\usepackage{bm}% bold math
\usepackage{amssymb}
\usepackage{amsmath}

\begin{document}
\title{Stochastic models for tumoral growth}

\author{Carlos Escudero}

\affiliation{Departamento de F\'{\i}sica Fundamental, Universidad Nacional de Educaci\'{o}n a Distancia,
C/ Senda del Rey 9, 28040 Madrid, Spain}

\begin{abstract}
Strong experimental evidence has indicated that tumor growth belongs to the molecular beam epitaxy universality class. This type of growth is characterized by the constraint of cell proliferation to the tumor border, and surface diffusion of cells at the growing edge. Tumor growth is thus conceived as a competition for space between the tumor and the host, and cell diffusion at the tumor border is an optimal strategy adopted for minimizing the pressure and helping tumor development. Two stochastic partial differential equations are introduced in this work in order to correctly model the physical properties of tumoral growth in (1+1) and (2+1) dimensions. The advantages of these models is that they reproduce the correct geometry of the tumor and are defined in terms of polar variables. Analysis of these models allow us to quantitatively estimate the response of the tumor to an unfavorable perturbation during the growth.
\end{abstract}

\maketitle

In Nature, one can find a huge number of systems that develop a rough interface in the process of growing. Many of them
have been adequately understood by means of the use of some tools from fractal geometry, as scaling analysis, as well as
via modelling with stochastic partial differential equations (SPDEs) and discrete models~\cite{barabasi}.
While these concepts do not constitute an exclusive theoretical framework for surfaces in the physical world, they can
be applied to win a deeper understanding of many processes in biology~\cite{eden}. Due to the many possible important applications in medicine, tumor growth constitutes one of the most interesting subjects of study to which scaling analysis can be applied. Actually, a very important research on tumor growth has been recently carried out. It has been found a strong empirical evidence that a broad class of tumors belong to the same universality class: the molecular beam epitaxy (MBE) universality class~\cite{bru1,bru2}. MBE dynamics are characterized by a number of features which include a linear growth rate, the constraint of growth activity to the outer border of the tumor, and surface diffusion at the growing edge; all of them have been observed experimentally. Surface diffusion has been identified as an optimal mechanism for favoring tumor growth. The host tissue exerts pressure on tumors which opposes their growth, but surface diffusion drives the cells to the concavities of the interface keeping this pressure to a minimum. These findings suggested studying the effect of the immune response on the tumor, and it has been established that an enhancement of the immune response increases the pressure on the tumor surface, and therefore limits its development~\cite{bru3}. A very important consequence of this fact is its possible application to improve cancer therapy, something that has been already exploited with positive results~\cite{bru4}. All these achivements underline the fundamental importance of understanding the physics of tumor growth.

Before introducing any model, it is of fundamental importance to point out that the application of a physical model to
a complex biological process implies a high simplification of many of its features. The movement of the cells is actually
much more complex than that simply described by diffusion, it is affected by chemotaxis and haptotaxis. The dynamics of
tumor-host interactions is determined by many complex cellular and extracellular processes, which include normal epithelial
and mesenchymal cells as well as the extracellular matrix in addition to the immune response. This interaction is very
complex and highly variable. Normal cells adjacent to the tumor are often induced to promote tumor growth by releasing
proteolytic enzymes to break down the extracellular matrix or release growth factors that enhance tumor proliferation.
In addition, it appears that tumor cells may under some circumstances transform into mesenchymal cells producing new
populations of relatively normal appearing cells that support tumor growth. Tumors are extensively infiltrated by
immune cells which may constitute as much as one third of its volume. Both the tumor phenotype and the tumor environment
are very heterogeneous. The former is the result of accumulating random mutations, variable enviromental selection forces
and perhaps restriction of proliferate capacity in non-stem cell components of the tumor. In addition, the tumor environment is extremely heterogeneous primarily due to disordered angiogenesis and blood flow. These facts underline the
huge complexity of the problem, and remind us that the equations appearing below do not constitute a fundamental
description of tumor growth, but a statistical approach to some of its properties.

The continuum equation which describes the MBE universality class (also known as the Mullins-Herring equation~\cite{mullins}) is the following SPDE
\begin{equation}
\partial_t h=-K\nabla^4 h + F + \eta({\bf x},t),
\end{equation}
where $h$ is the interface height, $K$ is the surface diffusion coefficient, and $\eta({\bf x},t)$ is a gaussian noise
with zero mean and correlations given by
$<\eta({\bf x},t)\eta({\bf x'},t')>=D\delta({\bf x}-{\bf x'})\delta(t-t')$.
The term $F$ has the dimensions of a velocity, and in this case should be interpreted as the product of the mean cell radius and the cell division rate. The critical exponents can be extracted from this equation simply by power counting.
If we neglect for a moment the velocity $F$, and we perform the transformations $x \to b x$, $t \to b^z t$, and $h \to b^\alpha h$ in the one-dimensional case, we see that the only values of $z$ and $\alpha$ that yield scale invariance are $z=4$ and $\alpha=3/2$. Since in this case $\alpha>1$, the system is super-rough and it is characterized by the set of critical exponents: $\alpha=3/2$, $\alpha_{loc}=1$, $z=4$, $\beta=3/8$, and $\beta^*=1/8$~\cite{lopez}, which were found to be compatible with those measured in experiments on tumor growth~\cite{bru1,bru2}. Another additional nice feature of the Mullins-Herring equation is its simplicity: since it is linear, it can be exactly solved by means of a Fourier transformation. Another properties of this equation, however, make it not so suitable for describing tumor growth. It describes the growth of a surface from a planar substrate of fixed size, while actual tumors show radial growth with their size continuously increasing in time. It is thus important to derive a SPDE able to describe MBE physics with the correct geometry and spatiotemporal properties of tumor growth.

In order to get the correct theoretical description of tumoral growth we will need to borrow some elements from differential geometry. On the other hand, these geometrical concepts are common in the formulation of stochastic growth equations in reparametrization invariance form~\cite{marsili}. The equation of growth of a general riemannian surface reads
\begin{equation}
\partial_t \vec{r}({\bf s},t)=\hat{n}({\bf s},t)\Gamma[\vec{r}({\bf s},t)]+\vec{\Phi}({\bf s},t),
\end{equation}
where the $d+1$ dimensional surface vector $\vec{r}({\bf s},t)=\{r_\alpha({\bf s},t)\}_{\alpha=1}^{d+1}$ runs over the surface as ${\bf s}=\{s^i\}_{i=1}^d$ varies in a parameter space (in the following, latin indices vary from 1 to $d$ and greek indices from 1 to $d+1$). In this equation $\hat{n}$ stands for the unitary vector normal at the surface at $\vec{r}$, $\Gamma$ contains a deterministic growth mechanism that causes growth along the normal $\hat{n}$ to the surface, and $\vec{\Phi}$ is a random force acting on the surface.
In our case the deterministic part should include a term modelling cell diffusion in the tumor border. When surface diffusion occurs to minimize the surface area the corresponding term in the equation is~\cite{marsili}:
\begin{equation}
\label{surdiff}
\Gamma_s=-K\Delta_{BL}H,
\end{equation}
where $\Delta_{BL}$ is the Beltrami-Laplace operator
\begin{equation}
\Delta_{BL}=\frac{1}{\sqrt{g}}\partial_i(\sqrt{g}g^{ij}\partial_j),
\end{equation}
$g_{ij}$ is the metric tensor and g is its determinant, $\partial_i=\partial/\partial s^i$ is a covariant derivative, and $H=\hat{n} \cdot \Delta_{BL} \vec{r}$ is the mean curvature.
Summation over repeated indices is always assumed along this work.
Finally, the unitary normal vector is given by
$\hat{n}=g^{-1/2}\partial_1\vec{r}\times\cdots\times\partial_d\vec{r}$.
In the case of the (1+1)-dimensional Monge form (or what is the same, the parametrization corresponding to a planar substrate) we have
$\vec{r}=(x,h(x))$, the unitary normal vector takes this times the form
\begin{equation}
\hat{n}=\frac{1}{\sqrt{1+(\partial_x h)^2}}(-\partial_x h,1),
\end{equation}
the metric tensor is given by (note that for this particular case the metric tensor is a scalar)
$\tilde{g}=1+(\partial_x h)^2$. Thus, the resulting mean curvature is
\begin{equation}
H=\frac{\partial_x^2 h}{[1+(\partial_x h)^2]^{3/2}}.
\end{equation}
The corresponding contribution to the drift reads
\begin{equation}
\Gamma_s=-K\frac{3[-1+5(\partial_x h)^2](\partial_x^2 h)^3-10[\partial_x h + (\partial_x h)^3]\partial_x^2 h \partial_x^3 h + [1+(\partial_x h)^2]^2\partial_x^4 h }{[1+(\partial_x h)^2]^{9/2}},
\end{equation}
that constitutes an expression far more complex than that of the Mullins-Herring equation.
However, we can linearize this expression about the derivatives of $h$ to get
$\Gamma_s=-K\partial^4_x h$,
recovering the familiar drift of the MBE equation. This is the so called small gradient expansion, and is valid if sharp changes in the interface are absent~\cite{marsili}.
The other contribution to the dynamics comes from particle input
$\Gamma_F=\hat{n}\cdot <\vec{F}>$,
where $\vec{F}$ is the flux of cells generated at the interface. If we suppose that cell generation is an isotropic process we find $<\vec{F}>=F\hat{n}$, which implies $\Gamma_F=F$. However, the input of new cells $\vec{F}$ is a random process, giving rise to a stochastic contribution to the dynamics under the form of a noise $\eta=\hat{n}\cdot\vec{\Phi}$, where $\vec{\Phi}=\vec{F}-<\vec{F}>$. In summary, this implies that the stochastic force fulfills $\vec{\Phi}=\hat{n}\eta$ and $<\eta>=0$. Rearranging all the terms we recover the one-dimensional Mullins-Herring equation for MBE growth:
\begin{equation}
\label{mbe}
\frac{\partial h}{\partial t}=-K\frac{\partial^4 h}{\partial x^4}+F+\eta(x,t),
\end{equation}
where the noise, $\eta(x,t)$, has been assumed to be gaussian with correlation given by $<\eta(x,t)\eta(x',t')>=D\delta(x-x')\delta(t-t')$.
The drift of this equation comes originally from Eq.(\ref{surdiff}), which expresses the "diffusion of the mean curvature"
of the surface. This corresponds, also, to a homogeneization of the pressure; let us show this fact as follows. Normal cells
adjacent to the tumor exert force against new born tumoral cells. Concavities are sourrended by a higher number of normal
cells than convexities, and thus feel more pressure. Tumoral cells move along the tumor edge driven by the surface forces,
what causes the effect of a diffusion: tumoral cells are redistributed from convexities to concavities (this fact will be shown explicitly below by means of linear stability analysis). The "equilibrium" distribution corresponds to the spherically symmetric form, which presumably implies the homogeneization of the pressure all along the tumor edge; this same form implies the minimization of the mean curvature of the surface. The specific form of the cuartic derivative in this
equation has been deduced phenomenologically~\cite{bru1,bru2}.

Now that we have identified the physical mechanisms that have led us to Eq.(\ref{mbe}), we are in position to derive SPDEs describing the same physics but with geometrical properties compatible with those of a tumor.
For the case of (1+1)-dimensional circular model in polar coordinates we have
$\vec{r}=(r(\theta)\cos(\theta),r(\theta)\sin(\theta))$, the unitary normal vector reads
\begin{equation}
\hat{n}=\frac{1}{\sqrt{r^2+(\partial_\theta r)^2}}(-r\sin(\theta),r\cos(\theta)),
\end{equation}
the corresponding metric tensor is given this time by
$\tilde{g}=r^2+(\partial_\theta r)^2$, and the mean curvature is
\begin{equation}
H=\frac{r^2+2(\partial_\theta r)^2-r\partial_\theta^2 r}{[r^2 + (\partial_\theta r)^2]^{3/2}}.
\end{equation}
We are now ready to derive the diffusive drift
\begin{equation}
\Gamma_s=-\frac{K}{r^4} \left( \frac{\partial^2 r}{\partial \theta^2} + \frac{\partial^4 r}{\partial \theta^4} \right),
\end{equation}
correspondingly linearized with respect to the different derivatives of $r(\theta)$. The term containing the second derivative of $r$ is irrelevant in the renormalization group sense, so we can neglect it to obtain
\begin{equation}
\Gamma_s=-\frac{K}{r^4} \frac{\partial^4 r}{\partial \theta^4}.
\end{equation}

The stochastic term comes from the force in the same way as in the last case $\vec{\Phi}=\hat{n}\eta$. The noise $\eta$ is a gaussian variable with zero mean and correlation given by
\begin{equation}
<\eta({\bf s},t)\eta({\bf s'},t')>=n_\alpha({\bf s},t)n_\beta({\bf s},t)D^{\alpha \beta}\frac{\delta({\bf s}-{\bf s'})} {\sqrt{g}}\delta(t-t').
\end{equation}
We now can write the SPDE for tumor growth in (1+1)-dimensions
\begin{equation}
\label{tumor2d}
\frac{\partial r}{\partial t}= -\frac{K}{r^4} \frac{\partial^4 r}{\partial \theta^4} + F + \frac{1}{\sqrt{r}}\eta(\theta,t),
\end{equation}
where the noise $\eta(\theta,t)$ is gaussian, with zero mean, and correlation given by
$<\eta(\theta,t)\eta(\theta',t')>=D\delta(\theta-\theta')\delta(t-t')$.
As indicated above, we have assumed that cell generation is isotropic, which implies $D^{\alpha \beta}=D\delta^{\alpha \beta}$.
It is important to note that this time the noise is multiplicative, and that it \emph{must} be interpreted according to
It\^o, since all the deterministic contributions to the drift have been already extracted~\cite{marsili}. Another desirable characteristic of this equation is that the variable $\theta$ only varies in $[0,2\pi]$ at any time, which represents an advantage with respect to using a different coordinate, as for instance the arc length. The arc length (a magnitude more similar to $x$ in Eq.(\ref{mbe})) vary in an interval which depends on time (because the tumor grows), what makes more difficult to study the scaling properties of the model.
We can determine the critical exponents by power counting. The arc lenght of a circumference is $l=r\theta$, and taking into account that it scales as $l \to bl$, we deduce that the angle scales as $\theta \to b^{1-\alpha}\theta$. The other two variables scale as $r \to b^\alpha r$ and $t \to b^z t$; direct substitution reveals that Eq.(\ref{tumor2d}) is in the MBE universality class. Of course, the range of validity of this equation assumes that the interface shows neliglible overhangs in the radial direction compared with the size of the tumor, a fact that has been observed experimentally in many cases~\cite{bru1}.

Now we deal with a nonlinear equation, in contrast to Eq.(\ref{mbe}), that cannot be solved simply by means of a Fourier transformation. We can instead employ different techniques in order to get some insight into it. For a big enough tumor we can approximate the first moment of the radius by the solution the mean-field version of Eq.(\ref{tumor2d}), i.e., neglecting the noise term. On the other hand, the condition of possibility for formulating a continuous equation as an adequate description of a tumor is that it is composed by a sufficiently high number of cells. So describing the tumor with a continuous equation is in the same order of approximation of considering the mean-field level for computing the first moment of the radius. The first step in the analysis is to note that the deterministic version of Eq.(\ref{tumor2d}) admits radially symmetric solutions of the form
$r(\theta,t)=R(t)=Ft+R_0$,
where $R_0$ is the radially symmetric initial condition. It is easy to show the linear stability of this solution by
substituting $r(\theta,t)=R(t)+\rho(\theta,t)$ in Eq.(\ref{tumor2d}) with $D=0$, where $\rho$ is a small perturbation. The resulting equation for $\rho$ is
\begin{equation}
\label{fourier2d}
\frac{\partial \rho}{\partial t}=\frac{-K}{(Ft+R_0)^4}\frac{\partial^4 \rho}{\partial \theta^4}.
\end{equation}
Since the function $\rho$ is $2\pi$-periodic in the $\theta$ variable we can express it exactly in terms of a Fourier series
\begin{equation}
\rho(\theta,t)=\sum_{n=-\infty}^{\infty}\rho_n(t)e^{in\theta},
\end{equation}
and by direct substitution in Eq.(\ref{fourier2d}) we see that the Fourier modes obey
\begin{equation}
\frac{d\rho_n}{dt}=\frac{-Kn^4}{(Ft+R_0)^4}\rho_n.
\end{equation}
We can integrate exactly this equation to obtain
\begin{equation}
\label{perturbation2d}
\rho_n(t)=\rho_n(t_0)\exp \left( \frac{-Kn^4}{3F} \left[\frac{1}{(R_0+Ft_0)^3}-\frac{1}{(R_0+Ft)^3} \right] \right),
\end{equation}
where
\begin{equation}
\rho_n(t_0)=\frac{1}{2\pi}\int_0^{2\pi}\rho(\theta,t_0)e^{-in\theta}d\theta,
\end{equation}
and thus we see that the perturbation decreases in time provided $t>t_0$~\cite{note1}. It is also important to note that Eq.(\ref{perturbation2d}) might be interpreted as the response of the tumor to an external perturbation. Stochastic generation of new cells drives the tumor away from the radially symmetric form, while surface diffusion tries to restore it; redistribution of cell density after radial symmetry breaking follows the law Eq.(\ref{perturbation2d}).

Our next step will be to derive the corresponding equation for the growth of a (2+1)-dimensional interface.
We can parametrize the two-dimensional surface by means of the vector
$\vec{r}=(r(\theta,\phi)\sin(\theta)\cos(\phi),r(\theta,\phi)\sin(\theta)\sin(\phi),r(\theta,\phi)\cos(\theta))$.
Implying that the metric tensor is
\begin{equation}
\tilde{g}=\left( \begin{array}{cc}
     r^2+(\partial_\theta r)^2  & \partial_\theta r \partial_\phi r \\
     \partial_\theta r \partial_\phi r & r^2 \sin^2(\theta)+(\partial_\phi r)^2 \end{array} \right).
\end{equation}
The mean curvature can be derived from the metric tensor, however, the expression is so cumbersome that it cannot be handled
with simplicity. We can instead linearize this expression about the different derivatives of $r$ to get
\begin{equation}
H=\frac{1}{r^2}\left(-2r+\frac{\partial_\theta r}{\tan(\theta)}+\partial_\theta^2 r+\frac{\partial_\phi^2 r}{\sin^2(\theta)}\right).
\end{equation}
Collecting the results we see that for the (2+1)-dimensional model for spherical growth the SPDE reads
\begin{equation}
\label{tumor3d}
\frac{\partial r}{\partial t}= -\frac{K}{r^4}\left(\frac{\partial^4 r}{\partial \theta^4}+
\frac{2}{\sin^2(\theta)}\frac{\partial^4 r}{\partial \theta^2 \partial \phi^2}+
\frac{1}{\sin^4(\theta)}\frac{\partial^4 r}{\partial \phi^4}\right)+ F +
\frac{1}{r\sqrt{|\sin(\theta)|}}\eta(\theta,\phi,t),
\end{equation}
where the noise $\eta(\theta,\phi,t)$ is gaussian, with zero mean, and correlation given by
$<\eta(\theta,\phi,t)\eta(\theta',\phi',t')>=D\delta(\theta-\theta')\delta(\phi-\phi')\delta(t-t')$,
and it \emph{must} be again interpreted according to It\^o. As happened with the (1+1)-model, we get the desirable characteristic that the variables $\theta,\phi \in [0,2\pi]$.
In this case we see again that there exists the radially symmetric solution for the deterministic version of Eq.(\ref{tumor3d}):
$r(\theta,\phi,t)=R(t)=Ft+R_0$,
where $R_0$ is the radially symmetric initial condition. We can analize its linear stability by substituting the solution
$r(\theta,\phi,t)=R(t)+\rho(\theta,\phi,t)$
in Eq.(\ref{tumor3d}) with $D=0$, where $\rho$ is a small perturbation $2\pi$-periodic in both variables $\theta$ and $\phi$, and thus can be represented in the form of a Fourier series
\begin{equation}
\rho(\theta,\phi,t)=\sum_{n,m=-\infty}^{\infty}\rho_{n,m}(t)e^{in\theta+im\phi}.
\end{equation}
The Fourier modes obey the ordinary differential equation
\begin{equation}
\frac{d\rho_{n,m}}{dt}=\frac{-K}{(Ft+R_0)^4}\left(n^4+\frac{8}{3}m^2n^2+\frac{8}{3}m^4\right)\rho_{n,m},
\end{equation}
that can be integrated to yield
\begin{equation}
\rho_{n,m}(t)=\rho_{n,m}(t_0)\exp \left(\frac{-K}{3F}\left[\frac{1}{(R_0+Ft_0)^3}-\frac{1}{(R_0+Ft)^3}\right]
\left[n^4+\frac{8}{3}m^2n^2+\frac{8}{3}m^4\right]\right),
\end{equation}
where
\begin{equation}
\rho_{n,m}(t_0)=\frac{1}{4\pi^2}\int_0^{2\pi}\int_0^{2\pi}\rho(\theta,\phi,t)e^{-in\theta-im\phi}d\theta d\phi,
\end{equation}
implying the stability of the solution provided $t>t_0$~\cite{note2}. This is, as in the former case, the law that imposes density redistribution after radial symmetry breaking due to stochastic generation of new cells.

In conclusion, we have derived the equations of growth of the (1+1)- and (2+1)-dimensional tumor interfaces containing the physics of MBE in the appropiate geometry. These equations provide us a description of the tumor in a coordinate system that is polar, and since all the coordinates are angles we have the additional advantage that they vary in intervals
that are independent of time. They correctly predict the constant velocity growth regime found experimentally during the
initial phase of growth,
and a linear stability analysis of radial solutions allowed us to quantitatively estimate the law of density distribution of new generated cells. However, latter stages of growth are characterized by certain decelaration of the growth rate;
this fact is not captured by the present model, and it will be studied in the future.
We have assumed all along this work that the tumor is composed of an enough large number of cells so that the hydrodynamic description by means of continuous equations makes sense. If we want to describe small tumors a kinetic approach to the problem becomes necessary, as for instance a master equation formulation. Master equation descriptions of growth models are already present in the literature~\cite{sos}, and may be adapted for the present case of a tumor. Furthermore, we can project the master equation for the cell population into a SPDE via field theoretic arguments~\cite{escudero} in order to recover a more similar theoretical approach to the one presented here. These and other questions will be the object of future research.

This work has been partially supported by the Ministerio de Ciencia y Tecnolog\'{\i}a (Spain) through Project No. BFM2001-0291 and by UNED.


\begin{thebibliography} {99}
\bibitem{note1} Except for the mode $n=0$, which is trivial since it implies perturbations homogeneous in $\theta$.
\bibitem{note2} Except for the mode $m=n=0$, which is trivial since it implies perturbations homogeneous in $\theta$ and $\phi$.
\end{thebibliography}
\end{document}